\documentclass[a4paper,11pt,openright]{elsarticle}      

\journal{Physica D}

\usepackage{amsmath,amssymb}
\usepackage{graphicx}
\usepackage{mathptmx}      
\usepackage{bm}

\def\Z{\mathbb{Z}}

\def\R{\mathcal{R}}
\def\E{\mathcal E}
\def\C{\mathcal C}
\def\p{\mathcal P}
\def\H{\mathcal H}
\def\M{\mathcal M}
\let\e=\epsilon
\let\d=\delta
\let\D=\Delta

\newcommand{\be}{\begin{equation}}    
\newcommand{\ee}{\end{equation}}
\newcommand{\ba}{\begin{eqnarray}}
\newcommand{\ea}{\end{eqnarray}}

\def\bbm[#1]{\mbox{\boldmath $#1$}}

\newtheorem{oss}{Remark}

\begin{document}

\begin{frontmatter}

\title{An energy-momentum map for the time-reversal symmetric 1:1 resonance with $\Z_2\times\Z_2$ symmetry}

\author{Giuseppe Pucacco\fnref{fn1}}
\fntext[fn1]{Corresponding author: pucacco@roma2.infn.it; telephone/fax: +39 06 72594541.}
\address{Dipartimento di Fisica and INFN -- Sezione di Roma II,
Universit\`a di Roma ``Tor Vergata", \\
Via della Ricerca Scientifica, 1 - 00133 Roma}

\author{Antonella Marchesiello\fnref{fn2}}
\address{Faculty of Nuclear Sciences and Physical Engineering,
Czech Technical University in Prague, D\v{e}\v{c}\'{\i}n Branch,
Pohranicn\'{\i} 1, 40501  D\v{e}\v{c}\'{\i}n}
\fntext[fn2]{anto.marchesiello@gmail.com}

\begin{abstract}
We present a general analysis of the bifurcation sequences of periodic orbits in general position of a family of reversible 1:1 resonant Hamiltonian normal forms invariant under $\Z_2\times\Z_2$ symmetry. The rich structure of these classical systems is investigated both with a singularity theory approach and geometric methods. The geometric approach readily allows to find an energy-momentum map describing the phase space structure of each member of the family and a catastrophe map that captures its global features. Quadrature formulas for the actions, periods and rotation number are also provided.

\end{abstract}


\end{frontmatter}

\section{Introduction}
\label{intro}

Among low-order resonances (see e.g.\cite{CDHS}) the Hamiltonian 1:1 resonance plays a prominent role. A huge amount of work has been devoted to this study leading to  advances that almost covered the subject. We recall the works of Kummer \cite{Ku}, Deprit and coworkers \cite{D1,DE,M1}, Cushman and coworkers \cite{CR}, Broer and coworkers \cite{Br1:1} and van der Meer \cite{vdM}. The general treatment of the non-symmetric 1:1 resonance seems to have been done by Cotter \cite{Cotter} in his PhD thesis. With motivations mainly coming from applied dynamics \cite{P09,MP11,MP13,STK,tv}, our study covers the most general case of a detuned 1:1--resonant normal form invariant under $\Z_2 \times \Z_2$ symmetry by considering its versal unfolding with three parameters plus detuning \cite{VU11}. Although the treatment in Kummer's work \cite{Ku} is general enough to accommodate for detuning-like terms, their analysis is not explicit neither in his work nor in the others cited above. Moreover, bifurcation sequences in terms of the distinguished parameter (the `energy'), which are useful when comparing with numerical or laboratory experiments, are not explicitly given in the available references. 

We exploit threshold values for bifurcations of periodic orbits as a latch to unlock the general structure of phase-space. The approach of the  paper is based on the use of a regular reduction \cite{CB,KE} dividing out the $\mathbb{S}^1$ symmetry of the normal form. The reduced Hamiltonian is invariant with respect to a second $\Z_2$ symmetry: we exploit a singular reduction introduced by Han{\ss}mann and Sommer \cite{HS} which allows us to divide out this symmetry. This trick provides an effective geometric strategy to understand how the phase-space structure is shaped by all possible combinations of the parameters. As a coronation of the geometric approach, a two-parameter combination (the `catastrophe' map, \cite{STK}) allows us to represent the general setting in a suitable 2-plane and all possible bifurcation sequences are clearly represented in the plane of the values of integrals of motion, the energy-momentum map, that can be plotted to get information on fractions of phase-space volume pertaining to each stable family. Quadrature formulas for the actions, periods and rotation number can also be obtained. 

The plan of the paper is the following: in Section \ref{NOFO} we introduce the normal form Hamiltonian, discuss its symmetries and the corresponding versal deformation; in Section \ref{GR} we study the generic bifurcation sequences of this class of systems; in Section \ref{em} we introduce an energy-momentum map; in Section \ref{AARN} we discuss methods to compute actions, periods and rotation number;  in Section \ref{Conclusions} we resume the results.

\section{The normal form and its versal deformation}
\label{NOFO}

On the manifold with symplectic structure $d p_1 \wedge d q_1 + d p_2 \wedge d q_2$, we consider the normal-form Hamiltonian \cite{Cic}
\be\label{K11}
K (\bm{p},\bm{q}) = \sum_{j=0}^N K_{2j},
\ee
with
\be\label{Hzero} K_0 = \frac12(p_1^2+p_2^2+q_1^2+q_2^2) \doteq \E \ee
and higher-order terms satisfying $\{K_0,K_{2j}\}=0, \; \forall j=1,...,N$.
$K (\bm{p},\bm{q})$ is assumed to be invariant under the $\Z_2\times\Z_2$ group $\Gamma=\{\rm{Id},S_1,S_2,S_1 \circ S_2\}$, 
where 
\begin{eqnarray}
S_1&:&(p_1,p_2,q_1,q_2)\rightarrow(-p_1,p_2,-q_1,q_2)\label{spatial_symmetry1}\\
S_2&:&(p_1,p_2,q_1,q_2)\rightarrow(p_1,-p_2,q_1,-q_2)\label{spatial_symmetry2}
\end{eqnarray}
and the time reversion symmetry $ (p_1,p_2,q_1,q_2)\rightarrow(-p_1,-p_2,q_1,q_2). $

$K$ is characterized by a set of `external' control parameters (to be distinguished from the `internal' parameters fixed by the dynamics) that we collectively denote with $\alpha_i^{(j)}$. They are certain non-linear combinations of the parameters of the original physical model.

At zero order the two natural parameters are the unperturbed frequencies. In the present setting we assume they are not far from unit ratio and, after a rescaling, we assume that the departure from exact 1:1 ratio is given by the `detuning' parameter $\d$ \cite{Henrard,Sch,Vf}. By introducing the action-angle variables of $K_0$ with the transformation
\be\label{Vaa}
q_{\ell}=\sqrt{2J_{\ell}}\cos\phi_{\ell},\;\;\;\;\;p_{\ell}=\sqrt{2J_{\ell}}\sin\phi_{\ell}, \quad \ell=1,2,
\ee
so that $ K_0 = \E = J_1 + J_2$, the first order term of the 1:1 resonant $\Gamma$-invariant normal form can be assumed to be \cite{SV,mho}
\be
K_2 = \d J_1 + \alpha_1 J_1^2+ \alpha_2 J_2^2+ \alpha_3 J_1 J_2 \left[2+\cos 2 (\phi_1-\phi_2)\right],\ee
where for simplicity we have suppressed the upper index in the first-order parameters $\alpha_i^{(1)}\doteq\alpha_i, i=1,2,3$.
In view of its peculiar role we include $\d$ in the category of internal (or `distinguished' parameters) \cite{Br1:1} and consider $\d J_1$ as a higher-order term with respect to $K_0$. We observe that the $\alpha_i^{(j)}$'s may in turn depend on $\d$ (as it happens, for example, in the family of natural systems with elliptical equipotentials \cite{MP13}). The higher-order terms $K_{2j}(J_{\ell},\phi_{\ell}),j>1,$ are homogeneous polynomials of degree $2j$ in $J_{\ell}$ depending on angles only through the combination $2 (\phi_1-\phi_2)$. One of the $\Z_2$ symmetries could be broken by adding one further external parameter \cite{MRS1,MRS2,P09}.

The canonical variables $J_{\ell},\phi_{\ell}$ are the most natural to investigate the dynamics in a perturbative framework. However, several other coordinate systems can be used to unveil the aspects of this class of systems. 
We list those that will be useful in the following. First of all we use  coordinates `adapted to the resonance' \cite{SV}. There are various ways to do this: in the following we exploit the canonical transformation \cite{Br1:2}
\be \label{suber}
\left\{
  \begin{array}{l}
    J_1=J \\
    J_2=\E-J \\
     \psi=\phi_2-\phi_1\\
    \chi=\phi_2.
  \end{array}
\right.
\ee
This is used to perform a first reduction of the normal form, since $\chi$ is cyclic and its conjugate action $\E$ is the additional integral of motion. To first order, the reduced Hamiltonian is
\be\label{RK}
{\mathcal K}_a = \E + \alpha_2\E^2+\left(\d - 2 (\alpha_2+\alpha_3)\E\right)J+(\alpha_1 +\alpha_2-2\alpha_3)J^2 + \alpha_3 J(\E-J) \cos 2 \psi.\ee
A further reduction into a planar system, viewing $\E$
as a \emph{distinguished} parameter \cite{Br1:2} is then obtained via the canonical transformation \cite{Ku}
\be\label{ccoord}
\left\{
  \begin{array}{l}
    x=\sqrt{2J}\cos\psi \\
    y=\sqrt{2J}\sin\psi.
  \end{array}
\right.
\ee
 In the subsequent section we work with these coordinates on which depends the universal deformation.
 
Following \cite{CB}, a different path to reduce the symmetry of the normal form passes through the introduction of the invariants of the isotropic harmonic oscillator:
\begin{equation}\label{invariants}
\left\{\begin{array}{ll}
         I_0= & \frac12(p_1^2+p_2^2+q_1^2+q_2^2)= K_0 =\E\\
         I_1= & p_1p_2+q_1q_2 \\
         I_2= & q_1p_2-q_2p_1 \\
         I_3= & \frac12(p_1^2-p_2^2+q_1^2-q_2^2).
       \end{array}
\right.
\end{equation}
The set $\{ I_0,I_1,I_2,I_3\}$ form a Hilbert basis of the ring of invariant polynomials and can be used as coordinates system for the reduced phase space. Their Poisson brackets are given by $\{I_a,I_b\}=2\e_{abc} I_c, \; a,b,c=1,2,3$. Notice that $I_0$ coincides with the linear part of the normal form $K_0=\E$, a Casimir of the Poisson structure. There is one relation between the new coordinates, namely $I_1^2+I_2^2+I_3^2=I_0^2=\E^2$, hence the sphere
\begin{equation}\label{phase_sphere}
\mathcal S=\left\{(I_1,I_2,I_3)\in\mathbb R^3\;:\;I_1^2+I_2^2+I_3^2=\E^2\right\}
\end{equation}
is invariant under the flow defined by \eqref{K11}. This provides a (geometric) second reduction to a one degree
of freedom system. The links between the two sets are given by the `Lissajous' relations
\cite{D1,DE}
\ba
I_1 &=& 2 \sqrt{J_1 J_2} \cos\psi = 2\sqrt{ J (\E-J)} \cos\psi,\label{inv1}\\
I_2 &=& 2 \sqrt{J_1 J_2} \sin\psi = 2 \sqrt{J (\E-J)} \sin\psi\label{inv2}
\ea
and
\be\label{stereo}
x=\frac{I_1}{\sqrt{\E-I_3}},\quad y=\frac{I_2}{\sqrt{\E-I_3}}.\ee
We remark that the coordinates $x,y$ are shown by Kummer \cite{Ku} to be associated with a variant of the stereographic projection of $\mathcal S$ on the $(I_1,I_2)$-plane.

The `normal modes' of the system are expressed in the following forms:
\be\label{NM1}
{\rm NM1,NM2}: \quad I_1 = I_2 = 0, \quad J=0,\E, \quad I_3 = \mp\E. \ee
The periodic orbits `in general position' are most simply derived from the fixed points of the Hamiltonian vector field associated with \eqref{RK}. The family of `inclined' periodic orbits corresponds to the in-phase oscillations
\be \label{Ia}
{\rm Ia,Ib}: \quad \psi= 0, \pi, \quad I_2 = 0, \quad I_3 = I_{3U}, \quad I_1 = \pm \sqrt{\E^2 - I_{3U}^2},
\ee
whereas the family of `loop' periodic orbits corresponds to the oscillations in quadrature
\be\label{La}
{\rm La,Lb}: \quad \psi= \pm \pi/2,  \quad I_1 = 0, \quad I_3 = I_{3L}, \quad I_2 = \pm \sqrt{\E^2 - I_{3L}^2}.
\ee
The expressions of $I_{3U}$ and $I_{3L}$ can be found by solving the conditions for the fixed points of the flow and will be recovered in Section \ref{GR} relying on geometric arguments. 

An important result in the framework of singularity theory is that of inducing a generic function, defined around a critical point and depending on several parameters, from a simple germ and deformation depending on a small set of derived parameters \cite{Br1:1,GB1,GB2,Mar,HDS,Hlibro}. In the present case, starting from the general setting introduced in \cite{Br1:1}, a versal deformation of the family of systems \eqref{K11} is obtained in  \cite{VU11}. The easiest way to perform this further normalization is by exploiting the planar reduction and use the stereographic coordinates \eqref{ccoord}. Let us consider the resulting normal form
\ba
{\mathcal K}_b(x,y; \E,\d,\alpha_i^{(j)}) = \E + K_2 (x,y; \E,\d,\alpha_i^{(1)}) + ... + K_{2N} (x,y; \E,\d,\alpha_i^{(N)}).
\ea
It can be shown \cite{Br1:1,VU11} that there exists a $\mathbb{Z}_2\times\mathbb{Z}_2$-equivariant transformation   which `induces' ${\mathcal K}_b$ from the function 
\be
 F(x,y,u_k)=\e_1x^4+(\mu+u_3)x^2y^2+\e_2y^4+u_1x^2+u_2y^2,\label{F_uni}
 \ee
 namely, there exists a diffeomorphism
 \be
\Phi: \mathbb R^2 \times \mathbb R^{m+2} \longrightarrow  \mathbb R^2\times \mathbb R^3,\quad
 (x,y,\E,\d,\alpha_i^{(j)}) \longmapsto  \left(x,y,u_k \right),\ee
where $m$ is the dimensionality of the external-parameter space, such that ${\mathcal K}_b = F \circ \Phi.$


The coefficients $u_k, k=1,2,3,$ depend on the internal $\E,\d$ and external $\alpha_i^{(j)}$ parameters and are constructed in an algorithmic way with an iterative process carried out up to order $N$. Explicit expressions for $N=2$ are computed in \cite{VU11}. The coefficients $\e_1,\mu,\e_2$ are otherwise determined by the leading-order terms `at the singularity' $\E=\d=0$ and are expressed as the discrete set of constants
\be\label{germ}
\mu=\frac{2(A-2C)}{\sqrt{|(A-3C)(A-C)|}}, \;\;
\e_1 = \frac{A-3C}{|A-3C|}, \;\; \e_2 = \frac{A-C}{|A-C|},
\ee
where 
\be\label{pard}
A\doteq\frac14 (\alpha_1+\alpha_2),\;\;\;
     B\doteq\frac12 (\alpha_1-\alpha_2),\;\;\;
C\doteq\frac14 \alpha_3.
\ee
The function $F(x,y)$ provides the phase portraits on either surfaces of section of the normal form as they are determined by varying the parameters. Quantitative predictions for bifurcations around the resonance are given by the series expansion of the $u$ coefficients in terms of the internal parameters. If we content ourselves with qualitative aspects, these predictions are already determined by their first order expressions
\be
u_1 = \frac{\Delta+(B-2(A-3C))\E}{\sqrt{|A-3C|}},
\quad
u_2 = \frac{\Delta+(B-2(A-C))\E}{\sqrt{|A-C|}},
\quad
u_3 = 0,
\ee
where $
\Delta\doteq \d/2.$
We remark that these qualitative aspects cannot change anymore by the addition of higher-order contributions: predictions  become only quantitatively more accurate by considering higher-order terms up to some optimal order \cite{gce,pbb}. The quartic terms of the function $F(x,y)$ (with $u_3=0$ and coefficients as in \eqref{germ}) compose the {\it germ} of this resonance and the quadratic terms give its {\it universal deformation}. Exploiting the transformation \eqref{stereo} in order to use the invariant polynomials as phase-space variables, we can therefore adopt the function
\begin{equation}\label{hamiltonian_inv}
{\mathcal K}_I(I_1,I_2,I_3; \E)=\left(1+\Delta\right)\E +(A+2C)\E^2 +(B\E+\Delta)I_3 + C (I_1^2-I_2^2) +(A-2C)I_3^2
\end{equation}
on the reduced phase space given by the sphere \eqref{phase_sphere} to study the general behavior of the family.
There is a certain degree of redundancy in the external parameters, however as we see below there is  no strict reason not to keep them all, so we perform a general analysis of \eqref{hamiltonian_inv} for arbitrary values of the external parameters $A,B,C$ and the internal parameters $\Delta$ and $\E$. 

\section{Geometric reduction}\label{GR}

\subsection{Reduced phase space}

The two reflection symmetries now turn into the reversing symmetries $I_1\rightarrow-I_1$
and $I_2\rightarrow-I_2$. Their composition $(I_1,I_2,I_3)\rightarrow(-I_1,-I_2,I_3)$ gives a (non-reversing)
discrete symmetry of \eqref{hamiltonian_inv}. We perform a further reduction introduced by Han{\ss}mann and Sommer \cite{HS} to explicitly divide out this symmetry. This is given by the transformation
\begin{equation}\label{tr_lem}
\left\{
  \begin{array}{ll}
    X= I_1^2-I_2^2 \\
    Y= 2I_1I_2 \\
    Z= I_3
  \end{array}
\right.
\end{equation}
which turns the sphere \eqref{phase_sphere} into the `lemon' space
\begin{equation}\label{lemon}
\mathcal L=\left\{ (X,Y,Z)\in\mathbb R^3\;:\;X^2+Y^2=\left(\E+Z\right)^2\left(\E-Z\right)^2\right\}
\end{equation}
with Poisson bracket
$$\{f,g\}\doteq\left(\nabla f\times\nabla g,\nabla L\right)$$
where $(.,.)$ denotes the inner product and 
$L\doteq X^2+Y^2-\left(\E+Z\right)^2\left(\E-Z\right)^2.$ 
The Hamiltonian becomes
\begin{equation}\label{hamiltonian_lem}
{\mathcal K}_I(X,Z)=\left(1+\Delta\right)\E +(A+2C)\E^2 + C X+(B\E+\Delta)Z +(A-2C)Z^2.
\end{equation}
The lemon surface is singular at the points $\mathcal Q_1\equiv\left(0,0,-\E\right)$
and $\mathcal Q_2\equiv\left(0,0,\E\right)$, therefore, whereas the first reduction leading to \eqref{hamiltonian_inv} is regular, the reduction of the discrete symmetry is singular \cite{CB}.

To simplify the following formulae we omit the constant term from \eqref{hamiltonian_lem} by introducing $\H \doteq {\mathcal K}_I - \left(1+\Delta\right)\E -(A+2C)\E^2 $. In this way we finally obtain
\begin{equation}\label{ham_lem_simply}
\H(X,Z)= C X+(B\E+\Delta)Z +(A-2C)Z^2.
\end{equation}
Each integral curve of the reduced system defined by \eqref{ham_lem_simply} is given by the intersection between $\mathcal L$ and the surface
\begin{equation} \label{Ham_surface}
\{(Z,X)\in\mathbb R^2\;:\;\mathcal \H=h\}
\end{equation}
and tangency points give equilibrium solutions. All information about bifurcations of periodic orbits in generic position and stability/transition of normal modes of the original system can be obtained by the study of the mutual positions of the surfaces $\H$ and $\mathcal L$  \cite{HS}. We can further simplify the approach by exploiting the fact that, since $Y$ does not enter in \eqref{ham_lem_simply}, the level sets $\{\H=h\}$ are parabolic cylinders. A tangent plane to $\mathcal L$ may coincide with a tangent plane to the parabolic cylinder $\{\mathcal K=h\}$
only at points where $Y$ vanishes: in order to study the existence and nature of the equilibria configuration of the system, it is then enough to  restrict the analysis to the phase-space section $\{Y=0\}$ .


For $A\neq2C\neq0$, if a tangency point occurs between  $\mathcal L$ and the surface \eqref{Ham_surface}, we have an (isolated) equilibrium for the reduced system.
Moreover,  two (degenerate) equilibria are represented by the singular points $\mathcal Q_1, \mathcal Q_2$. The contour $\mathcal C\equiv\mathcal L\cap\{Y=0\}$ in the $(Z,X)$-plane
is given by $\mathcal C_-\cup\mathcal C_+$, where
\be\label{lemon_arcs}
\C_\pm\equiv\left\{(Z,X)\in\mathbb R^2\;:\;|Z|\le\E,\;\,X=\pm\left(\E^2-Z^2\right)\right\}
\ee
and the set $\p\equiv\{\H=h\}\cap\{Y=0\}$ corresponds to the parabola
\begin{equation}\label{parabola}
X=\frac1{C} \left(h -(B\E+\Delta)Z-(A-2C)Z^2 \right)\doteq\p(Z).
\end{equation}
The reduced phase space $\mathcal L$ is invariant under reflection symmetries with respect to every coordinate-axes.
In particular, the reduced phase section $\C$ is invariant under both reflection transformations 
\be\label{R_z}
R_1:\; Z\rightarrow -Z, \quad
R_2:\; X\rightarrow -X
\ee
and their composition $R_2 \circ R_1$.  However the dynamics of the reduced system are not invariant under these actions.  Anyway it is easy to understand how they operate on the parabola $\eqref{parabola}$. When acting on $\p$, $R_1$ turns it into its symmetric with respect to the $X$-axis. Under the action of $  R_2$, $\p$ is reflected with respect to the $Z$-axis, that is, it reverses its concavity. Finally, the composition $ R_2  \circ R_1$ inverts the concavity of the parabola and then reflects it with respect to the $X$-axis (the application  of $ R_1\circ R_2$ on $\p$ gives the same result). Thus, we can restrict our analysis  to the case in which the parabola \eqref{parabola} is upward concave and for $\E=0$ achieves its minimum point on the negative $Z-$axis. If we choose a negative detuning, this corresponds to consider $A<2C$ and $C>0$. Here and in the following we refer to this case as the \emph{reference case}. Then, by a simple application of $R_1,R_2$ and/or their composition we obtain the bifurcation sequences in the remaining cases (cfr. the left panel in table \ref{T1}).

On the section $\C$, the two degenerate equilibria are $\mathcal Q_1\equiv(-\E,0)$ and $\mathcal Q_2\equiv(\E,0)$.
It is always possible to fix $h$ such that \eqref{parabola} intersects $\C$ in one of these points, so that
\ba
h&=&h_1\doteq \E \left((A - B - 2 C) \E - \Delta \right),\label{yax_energy}\\
h&=&h_2\doteq \E \left((A + B -2 C) \E + \Delta \right). \label{xax_energy}
\ea
 Thus, for $h=h_1$ the system stays in the point $\mathcal Q_1$
 and similarly for $h=h_2$. Comparing with \eqref{NM1} we see that they correspond to the two \emph{normal mode} solutions NM1 and NM2. A stability/instability transition of a normal mode is generally associated with the bifurcation of new periodic orbits. If this is the case, one or more tangency points arise between the reduced phase space section $\C_{\pm}$ and the parabola \eqref{parabola} .

\begin{table}
\centering
\begin{tabular}{c|ll}

                                  & $A<2C$& $A>2C$   \\ \hline
                                  &       &           \\
  $C>0$                           & $\mathcal I$ &$R_2\circ R_1$\\
  $C<0$                           & $ R_2$ &$ R_1$         \\
  \hline
\end{tabular}
  \qquad\qquad
\begin{tabular}{c|llll}

                                  & $\mathcal Q_1$ & $\mathcal Q_2$ &$ \mathcal Q_L$       & $ {\mathcal Q}_U$       \\ \hline
                                  &       &       &             &              \\
  $ R_1$                  & $\mathcal Q_2$ & $\mathcal Q_1$ &$\widetilde {\mathcal Q}_L$ & $\widetilde {\mathcal Q}_U$ \\
  $ R_2$                  & $\mathcal Q_1$ & $\mathcal Q_2$ &${\mathcal Q}_U$        & $ {\mathcal Q}_L$ \\
  $ R_2 \circ R_1$ & $\mathcal Q_2$ & $\mathcal Q_1$ &$\widetilde {\mathcal Q}_U$ & $\widetilde {\mathcal Q}_L$ \\
  \hline
\end{tabular}

\caption{\small{Starting from the reference case \ref{caso_particolare}, we obtain all the complementary cases using reflection symmetries $ R_1$, $ R_2$ and $ R_2 \circ R_1 $. $\mathcal I$ stands for the identity transformation. The right panel shows how the fixed points of the system change under the action of the reflection symmetries of the twice reduced phase space.}}
\label{T1}
\end{table}

\subsection{Reference case}\label{caso_particolare}

We start by introducing the following threshold values for $\E$:
\be\label{TE}
\E_{U1,2}\doteq \frac{\D}{\pm 2(A-3C)-B}, \quad 
\E_{L1,2}\doteq \frac{\D}{\pm 2(A-C)-B}
\ee
and observing that the parabola \eqref{parabola} has its vertex in
\begin{equation}\label{zm}
Z_V=\frac{B\E+\Delta}{2(2C-A)},\quad
X_V=\frac1{C}\left(h-\frac{(B\E+\Delta)^2}{4(2C-A)}\right).
\end{equation}
Therefore, in the case $A<2C$, $C>0$ and $\Delta<0$, the parabola is upward concave 
with a minimum in $Z_V$ which does not depend on $h$ and is negative for sufficiently small values of $\E$. The tangency points between $\p$ and $\C$ can be found by imposing that 
the discriminants of the quadratic equations 
\be\label{ZQZ}
\p(Z)=\pm \left(\E^2-Z^2\right)\ee
vanish. Accordingly, there is a tangency on $\C_+$,
\be\label{QU} 
{\mathcal Q}_U=\left(Z_U, \E^2-Z_U^2\right), \quad Z_U \doteq \frac{B\E+\Delta}{2(3C-A)},\ee
if
\begin{equation}\label{HU}
h=h_U\doteq C\E^2+\frac{(B\E+\Delta)^2}{4(3C-A)}=C\E^2+(3C-A)Z_U^2
\end{equation}
and a tangency on $\C_-$,
\be\label{QL} 
\mathcal Q_L=\left(Z_L, \E^2-Z_L^2\right), \quad Z_L \doteq \frac{B\E+\Delta}{2(C-A)},\ee
if
\begin{equation}\label{HL}
h=h_L\doteq -C\E^2+\frac{(B\E+\Delta)^2}{4(C-A)}=-C\E^2+(C-A)Z_L^2.
\end{equation}
Both solutions are subject to the constraints
\begin{equation}\label{U_existence}
-\E<Z_U,Z_L <\E.
\end{equation}
The first result \eqref{QU} determines a contact point on $\C_+$ for $\E>\E_{U1}$ if $2(A-3C)<B\leq2(3C-A)$ and for $\E_{U1}<\E<\E_{U2}$ if $B>2(3C-A)$. These bifurcations correspond to the two inclined orbits  \eqref{Ia} bifurcating {\it from} NM1 and annihilating {\it on} NM2. The nature of the fixed point can be assessed by computing its index \cite{Ku}: the contact point between $\p$ and $\C_+$ has index
\be\label{indU}
{\rm ind} ({\mathcal Q}_U) = {\rm sgn} [C(3C-A)].\ee
In the reference case, $C>0>A/2$ therefore ${\rm ind} ({\mathcal Q}_U) > 0$ and the inclined orbits are always stable.

On the lower branch, since it is necessary that $A\neq C$, in order to proceed we have to distinguish among the three sub-cases: 1. $A<C \quad (\e_1=\e_2=-1)$; 2. $C<A<2C \quad (\e_1=-1,\e_2=1)$; 3. $A=C$.

\subsubsection{$A<C$}\label{caso_particolare1}

In this sub-case the solution \eqref{QL} gives a tangency point $\mathcal Q_L$ on $\C_-$ for $\E>\E_{L1}$ if $2(A-C)<B\leq2(C-A)$ and for $\E_{L1}<\E<\E_{L2}$ if $B>2(C-A)$. The contact point between $\p$ and $\C_-$ has index
\be\label{indL1}
{\rm ind} (\mathcal Q_L) = {\rm sgn} [C(A-C)],\ee
therefore, in this sub-case, ${\rm ind} (\mathcal Q_L) < 0$ and loop orbits are unstable.


\subsubsection{$C<A<2C$}\label{caso_particolare2}

The existence and stability analysis of the system in sub-case $2$ follows almost the same way: however, the orbit structure turns out to be quite different since the concavity of the parabola is now smaller than that of the lower contour. 
If  $B\leq 2(A-3C)$ no contact points distinct from $\mathcal Q_1$ arise: as a consequence, the normal mode NM2 stays stable for all positive values of $\E$. If $2(A-3C)<B\leq 2(C-A)$, one contact point occurs  for $\E>\E_{U1}$ which corresponds to the bifurcation of the inclined orbits: they are stable as in the case before. If $2(C-A)<B\leq2(A-C)$, the conditions for tangency with the lower arc at $\mathcal Q_L$ are now satisfied for $\E>\E_{L2}$ and, if $B>2(A-C)$, for $\E_{L2}<\E<\E_{L1}$. 
The order of bifurcations is reversed and, since now ${\rm ind} (\mathcal Q_L) > 0$, loops are also themselves {\it stable}. 

The peculiarity of this sub-case is the `global bifurcation'. Let us consider the critical value of the distinguished parameter 
 \be \E_{GB} \doteq-\frac{\D}{B}.\label{engb}\ee
 Comparing with \eqref{QU}  and \eqref{QL}, we observe that
 \be 
 Z_U (\E_{GB}) = Z_L (\E_{GB}) = Z_V (\E_{GB}) = 0
 \ee
and we have a family of parabolas with axis coinciding with the $X$-axis. From \eqref{yax_energy}--\eqref{xax_energy}, at the value of the Hamiltonian
\be
h_1=h_2=h_{GB}=\frac{(A-2C)\D^2}{B^2},\ee
the parabola passes through \emph{both} points $\mathcal Q_1$ and $\mathcal Q_2$ and a simple computation shows that its minimum is negative but bigger than $\E^2$. For $h>h_{GB}$ we have stable inclined as before, for  $h<h_{GB}$ we have loops.


\subsubsection{Degenerate sub-case $A=C$} \label{proof_stat3}
If $A=C$, $\p$ and the lower arc of $\C$ have the same curvature. Hence, by a simple geometrical argument we see that if
$Z_V\neq0$, it is impossible to have any intersection point different from $\mathcal Q_1$ between $\p$ and $\C_-$.
Otherwise \emph{all} the points of the lower arc of $\C$ are tangency points between $\p$ and $\C$. Thus if $B>0$ $(B<0)$ and $\D<0$ $(\D>0)$, for $ \E=\E_{GB} $
we find infinite (non-isolated) equilibria given by all the points on $\C_-$. They correspond to the circle $I_1=0$ on the spherical reduced phase space \eqref{phase_sphere}. Only inclined orbits may bifurcate as isolated periodic orbits and this happens when a contact between $\p$ and $\C_+$ does occur. 

\begin{oss}
All cases with $\D>0$ can be treated as those with $\D<0$ by a transformation which exchanges the coordinate axes in the original phase space. On the reduced phase space  it corresponds to the reflection $R_1$. As a consequence, the equilibrium points $\mathcal Q_1$ and $\mathcal Q_2$ are exchanged and the parabola $\p$ is reflected into its symmetric with respect to the $X$-axis. 
\end{oss}

\subsection{Complementary cases}\label{compl}

In the previous section we considered the `reference' case $A<2C$ and $C>0$. Now we are going to study the dynamics of the system in the complementary cases: {\bf a)}: $A<2C$, $C<0$; {\bf b)}: $A>2C$, $C<0$; {\bf c)}: $A>2C$, $C>0$. As observed above, by applying the transformations \eqref{R_z} and their compositions, the orbital structure of the system in these cases can be deduced from the analysis of subsection \ref{caso_particolare}.

In case {\bf a)}, the critical value $Z_V$ does not change its sign, but the parabola $\p$ turns out to be downward concave.
However we can reverse its concavity by applying $R_2$. 
Since $R_2$ is a symmetry with respect to the $Z$-axis, the two degenerate equilibria are invariant under its action. On the other hand, if a tangency point occurs on $\C_+$ it is reflected into  a tangency point on $\C_-$ and vice-versa. This implies that the role of loop and inclined orbits is exchanged (cfr. the right panel in table \ref{T1}). Namely, the first periodic orbits to appear from NM1 are now the loop orbits. The corresponding threshold value for the distinguished parameter is again $\E=\E_{L1}$.
The bifurcation of inclined orbits is possible from NM1 in the case $A<3C$ for $\E>\E_{U1}$ and from NM2 for $\E>\E_{U2}$ in the case $3C<A<2C$. The degenerate case $A=3C$ is specular with respect to the case $A=C$ with $C>0$. It admits as an interesting example the family of natural systems with elliptical equipotentials \cite{MP13}: inclined are forbidden and only loop orbits may bifurcate as isolated periodic orbits when a contact between $\p$ and $\C_-$ occurs. 

In case {\bf b)}, $\p$ is upward concave and its maximum lies on the positive $Z$-axis. Thus, by applying $R_1$ we can deduce the orbital structure of the system from the case \ref{caso_particolare}. Under the action of $R_1$ the degenerate equilibria of the reduced system are exchanged. Furthermore, each tangency point between $\p$ and $\C$ is reflected into its symmetric with respect to the $X$-axis (cfr. the right panel in table \ref{T1}). Namely,
$$\mathcal Q_L\equiv(Z_L,X_L)\rightarrow\widetilde {\mathcal Q}_L\equiv(-Z_L,X_L),$$
$${\mathcal Q}_U\equiv(Z_U,X_U)\rightarrow\widetilde {\mathcal Q}_U\equiv(-Z_U,X_U).$$
Anyway, due to the singularity of the transformation \eqref{tr_lem}, to the points $\mathcal Q_L$ and $\widetilde {\mathcal Q}_L$ correspond the same two points on the section $I_1=0$ of the sphere \eqref{phase_sphere}, that is the same loop orbits for the two degree of freedom system. Thus loop orbits are invariant under the action of $R_1$. By a similar argument it follows the invariance of inclined orbits. However, since the degenerate equilibria on the reduced phase space are exchanged, if in the case \ref{caso_particolare}
a periodic orbit bifurcates from NM2, in case {\bf b)} it bifurcates from NM1 and vice-versa.

Finally, by applying $R_2 \circ R_1$ we obtain the stability analysis in case {\bf c)} from the case  \ref{caso_particolare}. The fixed points of the reduced system change according to the right panel of table \ref{T1}. As a consequence, the normal modes exchange their roles and the bifurcation order of inclined and loop orbits is reversed. 

%

\subsection{Degenerate cases}\label{dege}

There are two degenerate cases corresponding to the parameters values $C=0$ and $A=2C$. For $C=0$ the parabola $\p$ degenerates into a couple of straight lines both parallel to the $Z$-axis. Thus, for all positive values of $\E$, the system has only two equilibria represented by the singular points $\mathcal Q_1$ and $\mathcal Q_2$: the only periodic orbits allowed by the  two degree of freedom Hamiltonian are the normal modes.
This is not surprising since this case corresponds to two uncoupled non-linear oscillators.

In the case $A=2C$ and $C>0$, the parabola $\p$ degenerates into the straight line
\begin{equation}
X =h-\frac{B\E+\Delta}{C}Z.
\end{equation}
Let us denote it  by ${\mathcal Y}(Z)$. Its angular coefficient is given by
\begin{equation}
m\doteq-\frac{B\E+\Delta}{C}.
\end{equation}
For $\D<0$, $m$ is positive if and only if $B\leq0$ or $B>0$ and $\E<\E_{GB}$. Thus, for $\E<\E_{GB}$, if  $\mathcal Y$ passes through the point $\mathcal Q_1$, it may intersect the contour phase space $\C$ only in one further point  on its upper arc. The corresponding value for $h$ is given  by
\begin{equation}
h=\overline h:=-\frac{(B\E+\Delta)\E}{2C}.
\end{equation}
If this is the case, the fixed point $\mathcal Q_1$ results to be an unstable equilibrium. A similar argument shows that, if $m<0$ and $h=\bar h$, $\mathcal Y$ may intersect $\C$ only in one further point on its lower arc. Thus the critical value $\E=\E_{GB}$
does not determine a stability/instability transition for the fixed point $\mathcal Q_1$. As in the case $C<A<2C$, it corresponds to a global bifurcation for the system. In fact, for $m=0$, the straight line $\mathcal Y$ becomes parallel to the $X$-axis and for $h=0$ it passes through both degenerate fixed points. Hence, for $\E=\E_{GB}$ they turn out to be both unstable and their stable and unstable manifolds coincide.

Thus, the analysis of the nature of the normal mode NM1 for $\D<0$ gives that, if $-2C<B\leq 2C$, it becomes unstable for $\E>\E_{U1}$ and, for $B> 2C$, it is unstable for $\E_{U1}<\E<\E_{U2}$, where the thresholds are now given by
\be
\E_{U1,2} = -\frac{\D}{B\pm2C} .
\ee
By the symmetry of the reduced phase space,  if  $\mathcal Y$ intersect $\C$ on its upper arc for $h=\overline h$, then, by decreasing $h$ enough, it will intersect the contour phase space at $\mathcal Q_2$ and on one further point on $\C_-$. Thus, the fixed point $\mathcal Q_2$ turns out to be unstable exactly when also $\mathcal Q_1$ is! Indeed an easy computation shows that
$$\E_{U1}=\E_{L2}, \;\;\; \E_{L1}=\E_{U2}.$$
 Moreover, by the same argument used above, we see that a tangency point may occur on the upper arc of $\C$ if and only if a tangency point arises between $\mathcal Y$ and $\C_-$.
Hence   the fixed points $\mathcal Q_U$ and $\mathcal Q_L$ (and, as a consequence, loop and inclined orbits) bifurcate at the same time for $\E>\E_{U1}$ if $-2C<B\leq2C$ and for $\E_{U1}<\E<\E_{U2}$ if $B>2C$.

\begin{figure}
 \includegraphics[width=10cm]{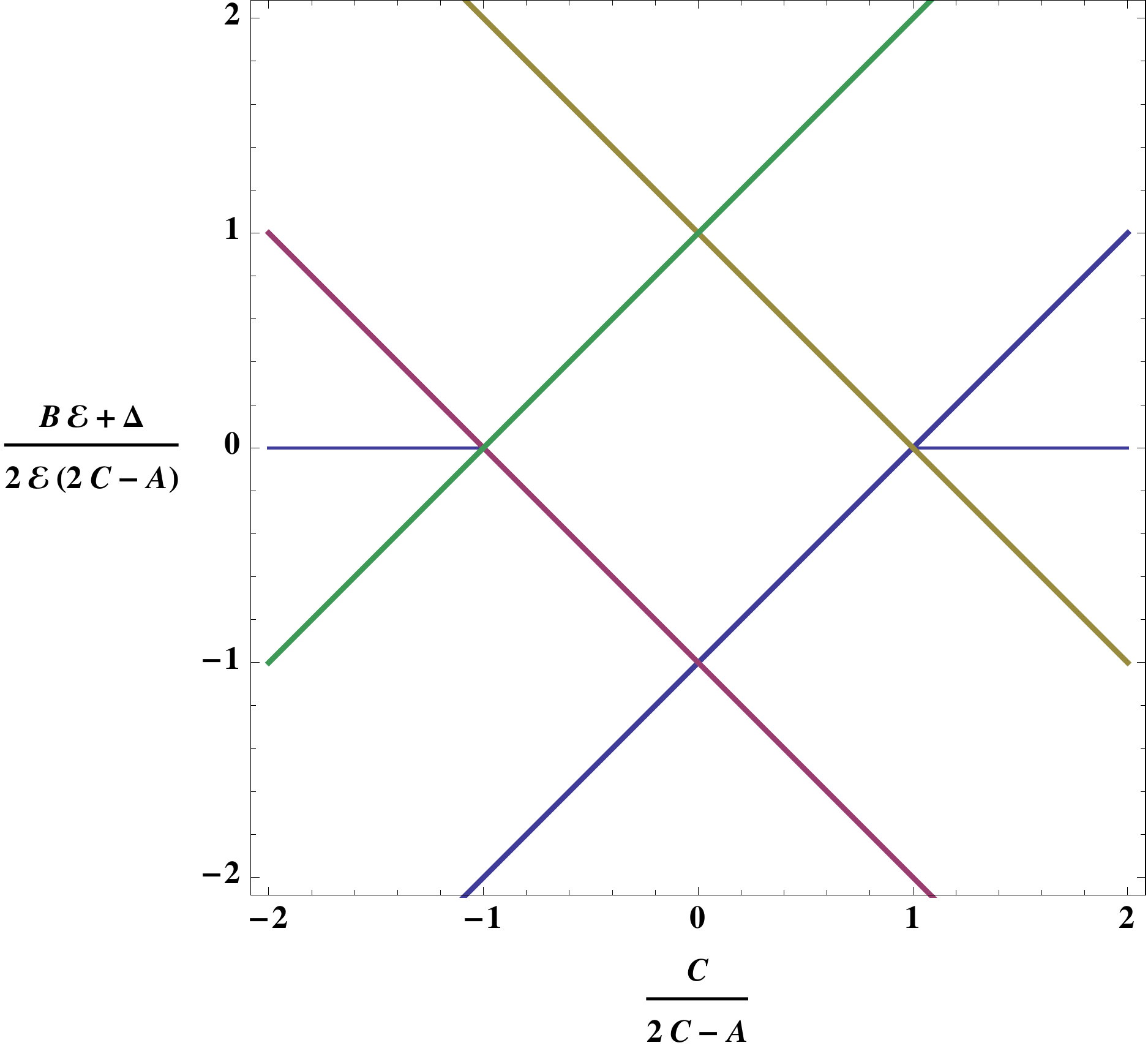}
\caption{Catastrophe map: the bifurcation lines are associated with $\E_{U1},\E_{U2}$ (eq.\eqref{zu1}, red and green lines), $\E_{L1},\E_{L2}$ (eq.\eqref{zl1}, blue and yellow lines).}
\label{cplot}       
\end{figure}

Since loop and inclined orbits bifurcate together, in the case $C<0$ the orbital structure of the system does not change, even if $\p$ reverses its concavity.

\subsection{Catastrophe map}\label{cata}

A comprehensive way to illustrate the general results described above is obtained by introducing  a pair of  combinations of the internal and external parameters and plot the bifurcation relations on the plane of this pair. This is referred to as the `catastrophe map' in the physical-chemical literature \cite{STK}. Recalling the four cases generated by the signs of $C$ and $2C-A$ (the reference and the complementary cases), we can use $C/(2C-A)$ 
as `coupling' parameter. A parameter which usefully combines the internal parameters $\E,\D$ with the remaining control parameter $B$ is the `asymmetry' parameter
\begin{equation}\label{zmp}
\frac{Z_V(\E)}{\E}=\frac{B\E+\Delta}{2(2C-A)\E}.
\end{equation}
By using the bifurcation values \eqref{TE}, we get
\ba
\frac{Z_V(\E_{U1,2})}{\E_{U1,2}}&=&\pm\frac{A-3C}{2C-A},\label{zu1}\\
\frac{Z_V(\E_{L1,2})}{\E_{L1,2}}&=&\pm\frac{A-C}{2C-A},\label{zl1}\ea
whereas the line
\begin{equation}\label{zmgb}
\frac{Z_V(\E_{GB})}{\E_{GB}}=0
\end{equation}
is associated with the global bifurcation. Plotting these lines on the plane of the coupling and asymmetry parameters (see fig.\ref{cplot}),  produces regions with no, one or two  families of periodic orbits in general position. The two triangular regions with bases on the lower/upper sides of the plot are below/above any bifurcation line, therefore they admit only normal modes. The central square is the locus with two bifurcations and therefore admits two families (one stable, the other unstable). The two triangular regions with bases on the lateral sides of the plot have two stable families: the horizontal segments are the loci of global bifurcation. The remaining regions have only one stable family of either type. 

\subsection{Physical application}\label{appl}

A physical interpretation of the classification obtained above concerns the relation between the phase-space structure and the strength of the nonlinear interaction between the two degrees of freedom. 

Considering the reference case and the complementary sub-case {\bf c)}, we have that, for $C>0$, if $C \in (A/3,A)$, both families of periodic orbits in generic position, if they exist, are stable; otherwise, one of the two families must be unstable. Recalling the definitions \eqref{pard} we deduce that, if the coupling  `physical' parameter $\alpha_3$ is such that
\be\label{interval}  (\alpha_1+\alpha_2)/3
   < \alpha_3 < \alpha_1+\alpha_2,\quad \alpha_3>0,\ee
   the system admits only stable bifurcating families.
   
   In the complementary sub-cases {\bf a)} and  {\bf b)}, it is straightforward to deduce that for $\alpha_3<0$ the system admits only stable bifurcating families if $\alpha_3$ stays in the complement of the interval defined by \eqref{interval}. In the light of application of singularity theory \cite{VU11}, the inclusion of small higher-order terms does not change these statements.

\section{An energy-momentum map for the $\Z_2\times\Z_2$ symmetric 1:1 resonance}\label{em}

The integrable dynamical system associated with the normal-form Hamiltonian \eqref{K11} gives the two-component map \cite{CB}
\ba
\E\M: && T^* \mathbb R^2 \longrightarrow  \mathbb R^2,\\
&& (p_1, p_2,q_1, q_2) \longmapsto  \left(K_0 (p_1, p_2,q_1, q_2),K (p_1, p_2,q_1, q_2)\right).\ea
The theorem of Liouville-Arnold \cite{Arn1} implies that, chosen a regular value $w$ of $\E\M$, there is a neighborhood $W(w)$ such that $\E\M^{-1} (W)$ is isomorphic to $W \times T^2$. This confirms that the phase-space of our system is a torus-bundle with (possible) singularities. By explicitly constructing the $\E\M$ map we can assess the nature of these singularities and how they are related with the critical values of the map. At critical values the differential of the energy-momentum map has rank less than two, therefore it is easy to guess that the curves of critical values on the image of the map are associated with the bifurcation lines found above and that the pre-image of the critical values coincide with the 1-tori of the periodic orbits in generic position \cite{CDHS,SD}.

For our purposes it is better to consider the map on the reduced phase-space. We have
\ba
\R\E\M: && \mathcal L \longrightarrow  \mathbb R^2,\\
&& (X,Y,Z) \longmapsto  \left(\E,\H \right).\ea

The rank of $d\R\E\M$ is zero at equilibrium and it is one where the differential of the two components are linearly dependent and not both zero. These conditions for the singular values of the map correspond to those exploited above in the geometric analysis. The $\H$ component assumes its extrema just on the normal modes and therefore the curves defined by \eqref{yax_energy}--\eqref{xax_energy} give the boundary branches of the image of the energy-momentum map up to the first bifurcation. The values of $\H$ at the contact points between the reduced phase-space and the second integral given by the functions \eqref{HU} and \eqref{HL} provide new branches starting and/or ending at bifurcating points. External branches are produced by stable bifurcations, the internal ones appear when unstable bifurcations are accompanied by the return to stability of a normal mode. All these features are nicely displayed in the bifurcation plots of the image of the map. 

\begin{figure}
 \includegraphics[width=12cm]{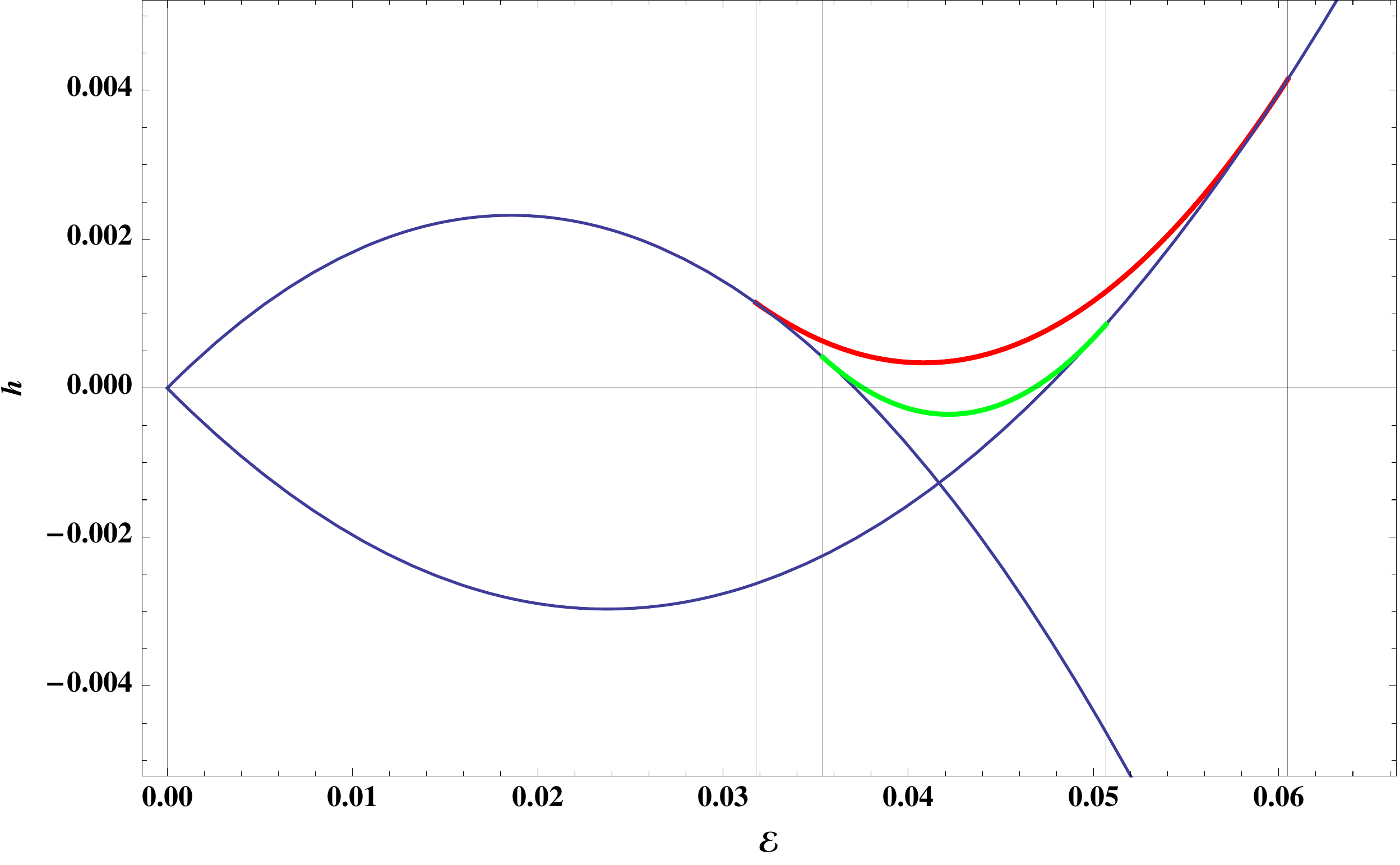}
\caption{Image of the $\E\M$ map in the case \ref{caso_particolare}, sub-case 1: $A=-1/3,B=6,C=1/5,\D=-1/4$.}
\label{rc1plot}       
\end{figure}

Let us consider for definiteness the reference case of subsection \ref{caso_particolare}. In fig.\ref{rc1plot} we see the image plot corresponding to the first sub-case, that with $A<C$: the vertical lines are given by the sequence $\E_{U1},\E_{L1},\E_{L2},\E_{U2}$ and the range of the map is the union of the 3 domains $\{0\le\E\le\E_{U1},h_2 \le h\le h_1 \}$, $\{\E_{U1}\le\E\le\E_{U2},{\rm min}(h_2,h_1)\le h \le h_U \}$ and $\{\E\ge\E_{U2},h_1 \le h\le h_2 \}.$ 
The thin blue curves correspond to the two normal modes. The red curve is associated with the bifurcation of the stable family of the inclined orbits whereas the green curve is associated with the bifurcation of the unstable family of the loop orbits: the `chamber' below it is occupied by invariant-tori around NM2 (again stable after $\E_{L1}$) which disappears when NM1 becomes unstable at $\E_{L2}$. 

\begin{figure}
 \includegraphics[width=12cm]{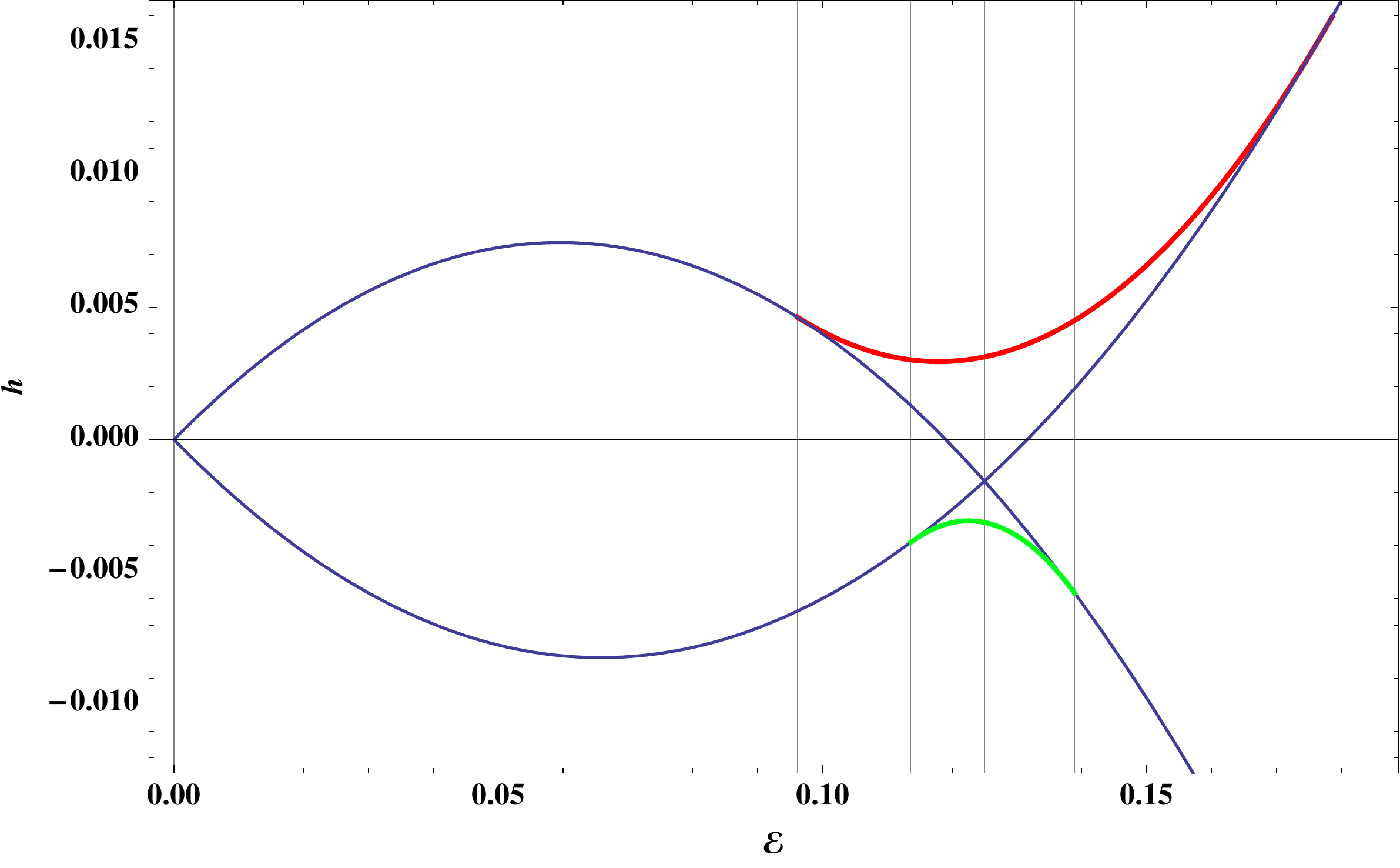}
\caption{Image of the $\E\M$ map in the case \ref{caso_particolare}, sub-case 2: $A=3/10,B=2,C=1/5,\D=-1/4$.}
\label{rc2plot}       
\end{figure}

In fig.\ref{rc2plot} we see the plot corresponding to the second sub-case, that with $C<A<2C$: the bifurcation sequence now is $\E_{U1},\E_{L2},\E_{GB},\E_{L1},\E_{U2}$ and the range of the map is the union of the 5 domains $\{0\le\E\le\E_{U1},h_2 \le h\le h_1 \}$, $\{\E_{U1}\le\E\le\E_{L2},h_2 \le h \le h_U \}$, $\{\E_{L2}\le\E\le\E_{L1},h_L \le h \le h_U \}$, $\{\E_{L1}\le\E\le\E_{U2},h_1 \le h \le h_U \}$ and $\{\E\ge\E_{U2},h_1 \le h\le h_2 \}$. The red curve is again associated with the bifurcation of the stable family of the inclined orbits whereas now the green curve is associated with the bifurcation of the {\it stable} family of the loop orbits and the chamber {\it above} it is occupied by invariant-tori parented by them. At the value $\E_{GB}$ corresponding to the global bifurcation the phase-space fraction of tori around the normal modes vanishes.

In both instances the parameters are chosen in order to have positive values for {\it all} the thresholds: otherwise, one or more branching points are lacking and the ensuing chambers are unbounded. The complementary cases of subsection \ref{compl} can be obtained by applying the transformation rules of table 1.

\section{Actions, periods and rotation number}
\label{AARN}

According to the Liouville-Arnold theorem \cite{Arn1} there exists a set of action-angle variables such that the Hamiltonian could be written in the form
\be\label{TAA}
{\mathcal K} = {\mathcal K} ({\mathcal J}_1,{\mathcal J}_2).\ee
The `frequencies' are accordingly found by means of the derivatives
\be\label{TFR}
\omega_{\ell}=\frac{\partial {\mathcal K}}{\partial {\mathcal J}_{\ell}}.\ee
The problem of finding expressions for the actions ${\mathcal J}_{\ell}, \ell=1,2,$ is simplified by the fact that $K_0=\E$ is already one of them, ${\mathcal J}_1\doteq\E$. The reduced dynamics investigated in the previous sections suggests to look for quadratures in $Z$. The canonical variables adapted to the resonance can be slightly modified with a linear transformation such that the symplectic structure becomes $d \E \wedge d \eta_+ + d Z \wedge d \eta_-$, with $\eta_{\pm} = (\phi_2\pm\phi_1)/2$. The second `non-trivial' action can therefore be computed by means of 
\be
{\mathcal J}_2 (\E,h)= - \frac1{2 \pi} \oint \eta_- dZ,\ee
where the contour of integration is the cross-section of the invariant torus fixed by $\E$ and $h$ on the $(Z,\eta_-)$-plane. By applying the linear transformation to the expressions of the invariants (\ref{inv1}--\ref{inv2}) and using the first of \eqref{tr_lem} we find
\be
\eta_-=\frac14 \arccos \left( \frac{X}{\E^2-Z^2} \right).\ee
The reduced dynamics is embodied in the relation \eqref{parabola} determining the parabola $X=\p(Z;\E,h)$. Henceforth, we obtain the following quadrature for the non-trivial action
\be\label{NTA}
{\mathcal J}_2 (\E,h)=- \frac1{8 \pi} \oint \arccos \left( \frac{\p(Z;\E,h)}{\E^2-Z^2} \right) dZ. \ee

With the approach adopted by Cushman and Bates \cite{CB} and successfully exploited in other resonant systems \cite{CDHS,TS,SD} we can express the `non-trivial' action by the linear combination
\be\label{NTAI}
{\mathcal J}_2= \frac1{2 \pi} T \left(\E,h \right)  {\mathcal K} - W \left(\E,h \right) {\mathcal J}_1.\ee
The two coefficients in the combination, depending only on the values of the integrals of motion, are respectively the {\it first return time}  $T$, or `reduced period' (divided by $2 \pi$), that is the time required to complete a cycle of the reduced Hamiltonian and the {\it rotation number} $W$ giving ($1/ 2 \pi \, \times$) the advance of the angle conjugate to the non-trivial action in a period $T$. These two statements can be proven by observing that, from ${\mathcal J}_1=\E$ and ${\mathcal J}_2= {\mathcal J}_2\left(\E,h \right)$ follows 
\be
\frac{\partial ({\mathcal J}_1,{\mathcal J}_2)}{\partial \left(h,\E \right)} = 
\left( \begin{array}{cc}
                        0 & 1 \\
                       \frac{\partial {\mathcal J}_2}{\partial h} & 
                       \frac{\partial {\mathcal J}_2}{\partial \E} \\
                      \end{array}
                    \right).\ee
 Then, in view of \eqref{TFR}, it can be readily proven \cite{TS} that 
 \be\label{PR}
 T =\frac{2 \pi}{\omega_2}=2 \pi \frac{\partial {\mathcal J}_2}{\partial h}\ee
 and
 \be\label{RN}
 W=\frac{\omega_1}{\omega_2}=-\frac{\partial {\mathcal J}_2}{\partial \E}.\ee
By using \eqref{NTA}, the reduced period \eqref{PR} is given by the quadrature
\be\label{QPR}
 T(\E,h) =\frac1{4C} \oint \frac{dZ}{\sqrt{Q(Z)}},\ee
 where we introduce the bi-quadratic 
 \be\label{Quartic}
 Q(Z;\E,h)=(\E^2-Z^2)^2-\left(\p(Z;\E,h)\right)^2.\ee
The rotation number is given by the partial derivative \eqref{RN}, being careful to recall the dependence of the reduced energy on $\E$:
\be\label{QRN}
 W(\E,h) =\frac1{8 \pi C} \oint \frac{(\E^2-Z^2)\left(1+\Delta+2(A+2C)\E + BZ\right) + 2 C \E \p(Z;\E,h)}{(\E^2-Z^2)\sqrt{Q(Z)}} dZ.\ee

These expressions are useful to assess general questions like monodromy, non-degeneracy conditions for the application of KAM theory, etc.\cite{CDHS,SD}. Here we exploit them to recover the frequencies of the periodic orbits. The integral \eqref{QPR} of the reduced period can be computed by extending to the complex plane and choosing a suitable contour determined by the roots of the polynomial $Q(Z)$. On periodic orbits we have double roots due to the tangency between the Hamiltonian and the reduced phase-space surfaces, therefore we obtain
\be\label{PROP}
 T(\E,h) =\frac14 \oint_{\gamma} \frac{dZ}{(Z-Z_C)\sqrt{a(Z-Z_1)(Z-Z_2)}},\ee
 where $Z_C$ is the contact point, $Z_{1,2}$ the other two roots of $Q(Z)=0$ and $\gamma$ is a cycle in the complex plane around the point $Z_C$. In the reference case, the constant $a$ is defined as
 \be
 a=(C-A)(3C-A);\ee
in the complementary cases a different choice of the sign can be necessary. Integrals of the form \eqref{PROP} can be computed with the method of residues. On the family of inclined, the double root is given by $Z_U$ in \eqref{QU}, so that
\be\label{PROPU}
 T_U \left(\E,h_U(\E) \right) = \frac{2 \pi i}{4C} {\rm Res} \left\{ \frac1{\sqrt{Q(Z_U)}} \right\}=
\frac{\pi}{2\sqrt{a(Z_{L1}-Z_U)(Z_U-Z_{L2})}},\ee
 where $Z_{L1,2}$ are the two distinct solutions of \eqref{ZQZ} evaluated at the reduced energy $h_U$ of \eqref{HU}. By explicitly computing the solutions and passing to the frequency we get
 \be\label{FNOPU}
 \omega_{2U}(\E) \doteq \frac{2 \pi}{T_U} =
2\sqrt{\frac{2C}{3C-A}}{\sqrt{\left( (2(A - 3C) - B)\E - \D\right)\left(\D + (2(A - 3C) + B)\E \right)}}.\ee
 Recalling the threshold values defined in the first of \eqref{TE}, we see that, in the reference case, the reduced frequency of inclined periodic orbits is real in their existence range $\E_{U1}\le\E\le\E_{U2}$, coherently with its interpretation as their normal frequency. 
 
 Proceeding in an analogous manner, with $Z_L$ double root of \eqref{ZQZ}, we get
 \be\label{PROPL}
 T_L \left(\E,h_L(\E) \right) = \frac{2 \pi i}{4C} {\rm Res} \left\{ \frac1{\sqrt{Q(Z_L)}} \right\}=
\frac{\pi}{2\sqrt{a(Z_{U1}-Z_L)(Z_L-Z_{U2})}},\ee
 where $Z_{U1,2}$ are the two distinct solutions of \eqref{ZQZ} evaluated at $h_L$ of \eqref{HL}. Accordingly
 \be\label{FNOPL}
 \omega_{2L}(\E) \doteq \frac{2 \pi}{T_L} =
2\sqrt{\frac{2C}{C-A}}{\sqrt{\left(\D - (2(A - C) - B)\E \right)\left(\D + (2(A - C) + B)\E \right)}}.\ee
From the threshold values defined in the second of \eqref{TE},  we again find that, in the reference case, we have to distinguish the two sub-cases $C>A$ and $C<A$: in the former, in the existence range $\E_{L2}\le\E\le\E_{L1}$, the argument of the square root is negative confirming the fact that the family of loops is unstable; in the latter, their reduced (normal) frequency is real and the family is stable. 
 
 We can use the quadrature for the rotation number to compute very easily the frequency $\omega_1$ of the periodic orbit itself. Let us denote for brevity with $A(Z)$ the argument of the integral in the expression \eqref{QRN}. On the family of inclined we obtain
 \be\label{RNOPU}
 W_U \left(\E,h_U(\E) \right) = \frac{i}{4C} {\rm Res} \left\{ A(Z_U) \right\}=
 \frac{i \left(1+\Delta+2(A+3C)\E + BZ_U \right)} {4C} {\rm Res} \left\{ \frac1{\sqrt{Q(Z_U)}} \right\},\ee
 from which, comparing with \eqref{PROPU}, we get
 \be\label{FROPU}
 \omega_{1U}(\E) = 1+\Delta+2(A+3C)\E + B \frac{B\E+\Delta}{2(3C-A)}.\ee
 Analogously, on the family of loops we have
 \be\label{RNOPL}
 W_L \left(\E,h_L(\E) \right) = \frac{i}{4C} {\rm Res} \left\{ A(Z_L) \right\}=
 \frac{i \left(1+\Delta+2(A+C)\E + BZ_L \right)} {4C} {\rm Res} \left\{ \frac1{\sqrt{Q(Z_L)}} \right\},\ee
from which, comparing with \eqref{PROPL}, we get
 \be\label{FROPL}
 \omega_{1L}(\E) = 1+\Delta+2(A+C)\E + B \frac{B\E+\Delta}{2(C-A)}.\ee

\section{Conclusions}
\label{Conclusions}

We have presented a general analysis of the bifurcation sequences of 1:1 resonant Hamiltonian normal forms invariant under $\Z_2\times\Z_2$ symmetry. The family of Hamiltonians is in a standard form of a universal deformation obtained from a singularity theory approach. The rich structure of these systems has been investigated with geometric methods. The bifurcation sequences of periodic orbits in general position are established by first reducing the normal form and than analyzing the relative equilibria by studying the intersection of the surfaces of the Hamiltonian and the twice reduced phase space. 

A generic exploration of the space of external control parameters is possible by first examining a reference set and then analyzing its complement by exploiting the symmetries of the system. An overall picture is provided by the reduced energy-momentum map for each inequivalent cases specified by the internal parameters. A global picture combining internal and external parameters is provided by plotting the catastrophe map. Finally, quadrature formulas for actions, periods and rotation number have been obtained. 

\section*{Acknowledgments}

We acknowledge useful discussions with H. Han{\ss}mann, G. Gaeta and F. Verhulst. G.P. is supported by INFN, Sezione di Roma Tor Vergata.

\end{document}